# Studying the Internal Ballistics of a Combustion Driven Potato Cannon using High-speed Video

E.D.S. Courtney[1] and M.W. Courtney[2]

[1] BTG Research, P.O. Box 62541, Colorado Springs, CO, 80962
[2] United States Air Force Academy,* 2354 Fairchild Drive, USAF Academy, CO, 80840
Michael_Courtney@alum.mit.edu

**Abstract**
A potato cannon was designed to accommodate several different experimental propellants and have a transparent barrel so the movement of the projectile could be recorded on high-speed video (at 2000 frames per second). Both combustion chamber and barrel were made of polyvinyl chloride (PVC). Five experimental propellants were tested: propane ($C_3H_8$), acetylene ($C_2H_2$), ethanol ($C_2H_6O$), methanol ($CH_4O$), and butane ($C_4H_{10}$). The amount of each experimental propellant was calculated to approximate a stoichometric mixture and considering the Upper Flammability Limit (UFL) and the Lower Flammability Limit (LFL), which in turn were affected by the volume of the combustion chamber. Cylindrical projectiles were cut from raw potatoes so that there was an airtight fit, and each weighed 50 (± 0.5) grams. For each trial, position as a function of time was determined via frame by frame analysis. Five trials were taken for each experimental propellant and the results analyzed to compute velocity and acceleration as functions of time. Additional quantities including force on the potato and the pressure applied to the potato were also computed. For each experimental propellant, average velocity vs. barrel position curves were plotted. The most effective experimental propellant was defined as the one which accelerated the potato to the highest muzzle velocity. The experimental propellant acetylene performed the best on average (138.1 m/s), followed by methanol (48.2 m/s), butane (34.6 m/s), ethanol (33.3 m/s), and propane (27.9 m/s), respectively.

**Introduction**
The potato cannon is a relatively small scale projectile launcher often used for physics demonstrations, projectile launching experiments and recreation. Pneumatic, or compressed air, potato cannons utilize air pressure to accelerate the projectile. In combustion driven potato cannons the expanding gases from the burning experimental propellant accelerate the projectile. Because they are simpler and less expensive to build than pneumatic potato cannons (e.g. Gurstelle 2001), combustion-driven potato cannons are common.

The internal ballistics of pneumatic potato cannons has been described by Mungan (2009), who showed that performance as measured by muzzle velocity was reasonable compared to predictions based on the laws of thermodynamics. Jasperson and Pollman (2011) described how muzzle velocity could be maximized by maximizing the product of the initial pressure and the volume of the experimental propellant gas and decreasing the projectile mass. Rohrbach et al. (2011) reported that predictions of muzzle velocity based on assumption of either purely adiabatic or isothermal process were inaccurate. They found better agreement with a model based on the flow of air through the valve. Although the internal ballistics of pneumatic potato cannons has been widely reported, the internal ballistics of the

---

combustion driven potato cannon have not been quantitatively described in detail.

In combustion driven potato cannons, the combustion chamber is filled with a propellant. Ignition may be achieved in several ways; such as a lantern sparker, a model rocket igniter, or an electric switch. As the propellant burns, the reaction creates rapidly expanding gases, and the pressure from the gases forces the projectile down the barrel. Many different propellants are possible, but aerosol hair spray is a common one for recreational and high school applications.

Performance of potato cannons is most commonly described using muzzle velocity. Jasperson and Pollman (2011) verified their predictions over a range of initial gauge pressures using high-speed video. Courtney and Courtney (2007) described an acoustic method for measuring muzzle velocity that is more widely accessible because it only requires a personal computer and free software. In previous studies, the internal ballistics of potato cannons was inferred from the muzzle velocity and not directly observed.

In this study, a potato cannon with a transparent barrel was designed so that the internal ballistics could be directly quantified using high-speed video. Several experimental propellants were investigated so that the general properties of the internal ballistics could be better understood. The transparent barrel allows velocity to be determined at any distance down the barrel rather than only at the muzzle.

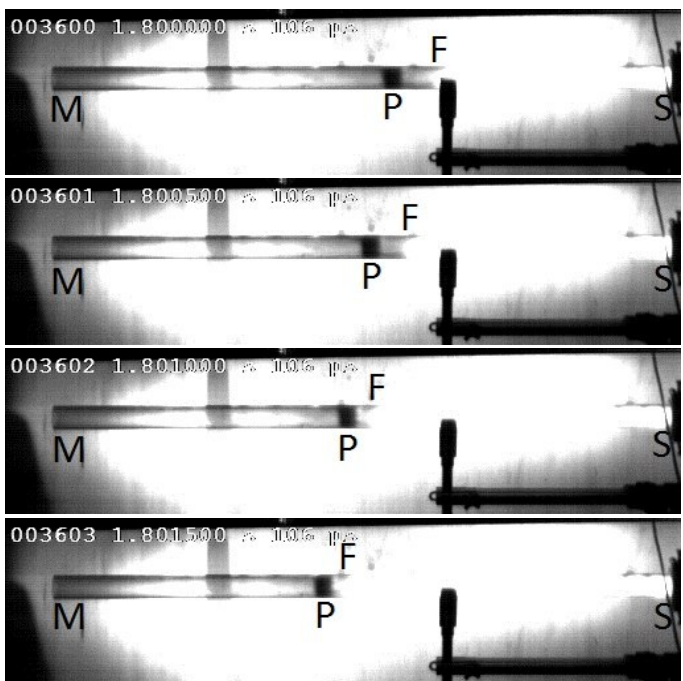

Figure 1. Sequence of video frames showing movement of potato from right to left in transparent barrel. P: potato, F: flame front, M: muzzle, S: starting position of potato. The chamber is barely visible on the right end of the frame.

**Method**

A combustion-driven potato cannon was designed and used so that the internal ballistics could be directly quantified using a high-speed video camera. The potato cannon's combustion chamber was a 45.7 cm long piece of PVC pipe, 6.4 cm inner diameter, and 1469.4 cubic centimeters in volume. The transparent PVC barrel was 121.9 cm long and 4.0 cm inner diameter. The experimental propellants were propane ($C_3H_8$), acetylene ($C_2H_2$), ethanol ($C_2H_6O$), methanol ($CH_4O$), and butane ($C_4H_{10}$). The amount of propellant transferred into the combustion chamber was determined using the ratio of the amount of available oxygen to the amount of fuel (propellant). Enough propellant was added to have at least a stoichiometric mixture without exceeding the upper flammability limit (Table 1).

Specified amounts of butane, propane, and acetylene (see Table 1) were measured in a 400 ml dental syringe and transferred into the combustion chamber through a small hole, which was then stopped with putty. For each trial of methanol and ethanol, a paper napkin was soaked in the respective propellant, placed in the combustion chamber, and left for approximately 90 seconds. In this manner, the amount of vaporized propellant in the combustion chamber was determined by the vapor pressure. Prior to ignition, a 50 gram (±0.5 grams) cylindrical piece of potato (*Solanum tuberosum)* was placed in the barrel. This projectile prevented the propellants from dissipating throughout the barrel before ignition. Five trials were conducted for each of 5 experimental propellants tested.

Due care should always be exercised when handling flammable gases and volatile liquids (Furr 2010). Safety procedures vary with institutional and regional requirements, but always include working in a well ventilated area (such as outdoors or under a fume hood) and keeping the work area free from possible ignition sources (including cigarettes, Bunsen burners, pilot lights, and flames).

To ignite the propellant, an electric match inserted through the ignition hole was triggered remotely using a DC source. The high-speed video camera (IDT Motion Pro X4, Integrated Design Tools, Pasadena, CA) was triggered manually at the same time the experimental propellant was ignited. Video data were recorded at 2,000 fps for two seconds. A frame-by-frame analysis of each video was performed using MotionStudio software (Integrated Design Tools, Pasedena, CA). A short sequence of frames is shown in Figure 1. The pixel position of the leading edge of the projectile in each frame was recorded in a spreadsheet. The length scale for the video data was determined using the known length of the barrel divided by the number of pixels between the beginning and the end of the barrel. The velocity of the projectile vs. time was calculated as the change in position divided by the time between frames. A polynomial model was fit to the velocity data using Graph.exe (http://www.padowan.dk/download/). The model was used to generate a smooth set of velocity vs. time data for further analysis.

Using the polynomial model, acceleration vs. time was computed as the change of velocity divided by the time between frames. Also using the polynomial model, the position of the potato in the barrel vs. time was computed. Each position was calculated as the sum of the changes in velocity times the time between frames up to that point.

The internal pressure created by the burning propellants is of general interest in internal balllistics, and also important to know to ensure the device does not rupture due to overpressure. The pressure was determined by dividing the estimated force on the projectile by the cross-sectional area. The force was estimated two ways to check for consistency. The work-energy theorem of physics was applied to estimate the force as the change in energy of the projectile divided by the change in position. The force estimated using the work energy theorem was divided by the cross-sectional area of the barrel to determine pressure as a function of barrel position. The force was also estimated as the mass of the projectile times its acceleration, which had been computed using the polynomial model. These latter force estimates were within a few percent of the former.

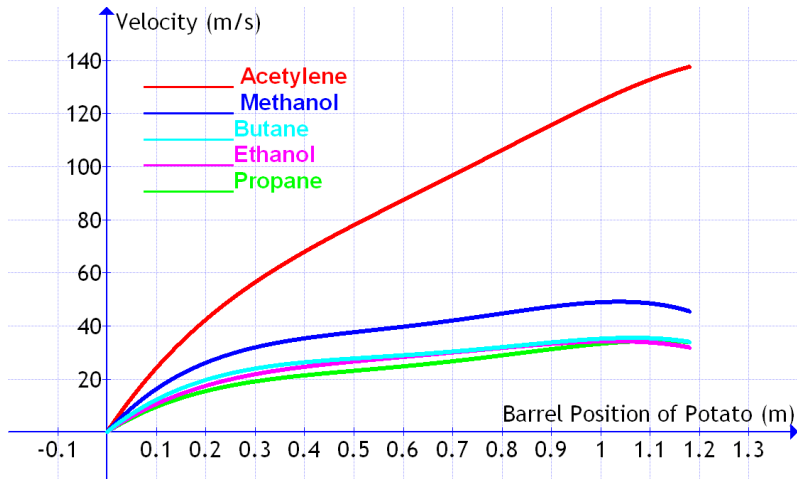

*Figure 2. Average velocity of cylindrical potato projectiles vs. barrel position for each experimental propellant.*

### Results

The results of this experiment showed that acetylene outperformed all of the other propellants by a large margin with an average muzzle velocity of 138.1 m/s (Figure 2). Methanol performed second most effectively with an average muzzle velocity of 48.2 m/s. Butane, ethanol, and propane had average muzzle velocities of 34.7 m/s, 33.3 m/s, and 27.9 m/s, respectively (Table 1). Velocities at 0.5 meters down the barrel were correlated with and in the same order as the muzzle velocities. The uncertainty in muzzle velocities varied from about 4% to about 18%, depending on the propellant.

| Propellant | Muzzle Velocity (m/s) | Uncertainty (%) | Velocity at 0.5 m (m/s) | Uncertainty (%) | Peak Pressure (kPa) | Amount of Propellant (cc) |
|---|---|---|---|---|---|---|
| Acetylene | 138.1 | 3.9 | 78.8 | 6.8 | 601.9 | 400 |
| Propane | 27.9 | 14.9 | 16.5 | 14.4 | 45.7 | 208 |
| Ethanol | 33.3 | 12.5 | 27.9 | 10.9 | 49.1 | 216 |
| Methanol | 48.2 | 17.9 | 38.1 | 15.3 | 127.2 | 96 |
| Butane | 34.7 | 15.1 | 28 | 14.2 | 62.5 | 160 |

*Table 1. The average muzzle velocity of a cylindrical potato projectile (five trials each) produced by each experimental propellant. The peak pressure computed for any trial for a given experimental propellant is reported for comparison with the peak pressure limit of 1720 kPa for the driving section, suggested by the manufacturer.*

It is of general interest in internal ballistics to know the pressure curves generated by each propellant. Figure 3 shows the pressure in the barrel (averaged over five trials) produced by each fuel as a function of the barrel position of the potato. For acetylene, barrel pressure increased sharply to about 300 kPa as the potato traveled the first 0.3 m of the barrel. Pressure continued to increase at an inconsistent but somewhat slower rate until reaching an average peak of about 430 kPa at about 1.05 m. The pressure was still more than 375 kPa when the potato exited the barrel but the pressure was declining. For each of the other four propellants, the pressure rose quickly to a peak as the potato traveled the first 0.23 meters of the barrel. The pressures then decreased until a barrel length of 0.7 m, plateauing until a distance of 0.95 m, before slightly increasing, and finally becoming negative at a barrel length of about 1.15 meters.

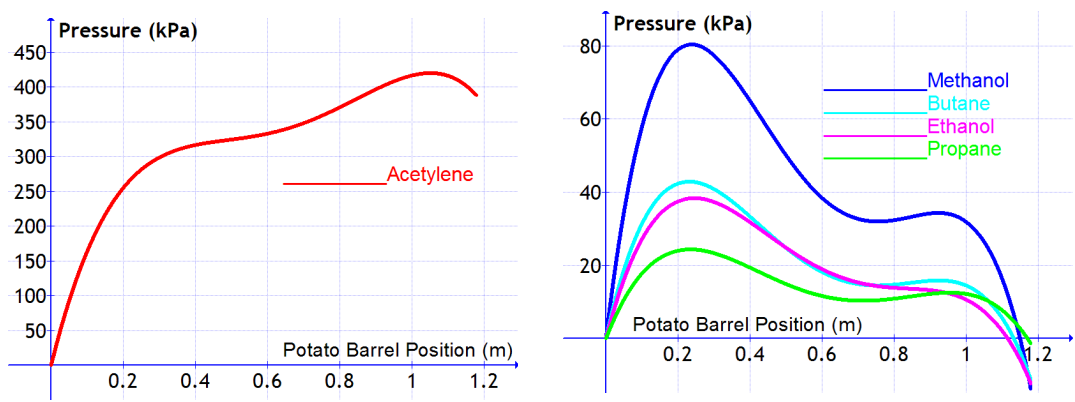

*Figure 3. Left) Average pressure curve of five trials produced by acetylene. Right) Average pressure curves of five trials produced by methanol, ethanol, butane, and propane.*

The peak pressures produced by the experimental propellants corresponded to the order of muzzle velocities. The manufacturer's upper pressure limit of the PVC used for the combustion chamber was 1720 kPa. The peak pressure of any of the five trials of acetylene was 601 kPa, just over a third of this limit. The other propellants produced pressures that were ten percent or less of the limit. This result suggests that this design will not fail due to excessive pressure under the conditions used in this experiment.

**Discussion**
The purpose of the experiment was to quantify velocities and internal pressures of a potato cannon using different propellants and a high-speed video camera. The hypothesis that acetylene would accelerate the projectile to the highest velocity was confirmed. For every propellant tested except acetylene, pressures became negative before the projectile left the barrel, (Figure 3) thus decreasing the projectile's velocity. However, if a shorter barrel had been used, these propellants could have reached maximum velocity at the muzzle. If a longer barrel had been used, acetylene would have kept accelerating the potato for some additional distance, before finally also becoming subject to negative pressures. The rank ordering of the average pressures generated by each propellant corresponded to the rank ordering of the average speeds for each propellant.

The amount of each propellant was between the amount needed for a stoichiometric mixture and the upper flammability limit, with the exception of ethanol. Ethanol was not used in this way because it would have required a more complex process to prepare and transfer the amount of vaporized ethanol needed for a stoichiometric mixture. At the outset, the experiment was designed to be practical and accessible for high-school or home-based applications. However, this design constraint introduced the potential limitation that the differences in performance were due to different quantities of propellants. To investigate this further, the efficiency of each propellant was computed by dividing the kinetic energy of the projectile at the muzzle by the chemical energy available for each propellant (Table 2).

The rank order of efficiencies corresponds to the muzzle and mid-barrel velocities of the projectile.

| Propellant | Combustion Energy (J) | Projectile Muzzle Energy (J) | Efficiency (%) |
|---|---|---|---|
| Acetylene | 10894 | 513 | 4.71 |
| Propane | 9302 | 21 | 0.23 |
| Ethanol | 9078 | 29 | 0.32 |
| Methanol | 10140 | 66 | 0.65 |
| Butane | 9273 | 33 | 0.35 |

*Table 2. Combustion energy, muzzle energy, and efficiencies of experimental propellants.*

Careless handling of a potato cannon could cause serious injury or death (Frank 2012). Potatoes launched with acetylene were also destructive to wooden boards and plastic objects initially employed as backstops before transitioning to 6mm thick steel plate. Adult supervision and due care regarding safe firing directions is imperative when using these devices.

**Acknowledgments**